\documentclass{article}
\usepackage{jheppub}

\usepackage{amsfonts}

\def\bea{\begin{eqnarray}}
\def\eea{\end{eqnarray}}
\def\nn{\nonumber}

\def\lmatrix{\left(\begin{array}}
\def\rmatrix{\end{array}\right)}
\def\gsim{\mathrel{\rlap{\lower4pt\hbox{\hskip1pt$\sim$}}\raise1pt\hbox{$>$}}}
\def\lsim{\mathrel{\rlap{\lower4pt\hbox{\hskip1pt$\sim$}}\raise1pt\hbox{$<$}}}
\def\bi{\begin{itemize}}
\def\ei{\end{itemize}}
\def\msbar{\overline{\rm MS\kern-0.5pt}\kern0.5pt}
\def\rho{\varrho}

\title{The semi-classical approximation at high temperature revisited}

\author{Alexander Boccaletti, }
\author{Daniel Nogradi}

\affiliation{Eotvos University, Department of Theoretical Physics, Pazmany Peter setany 1/a, Budapest 1117, Hungary}

\emailAdd{boccalex@caesar.elte.hu}
\emailAdd{nogradi@bodri.elte.hu}

\abstract{
We revisit the semi-classical calculation of the size distribution of instantons at finite temperature in non-abelian
gauge theories in four dimensions.
The relevant functional determinants were first calculated in the seminal work of Gross, Pisarski and Yaffe
and the results were used for a wide variety of applications including axions most recently. In this work we
show that the uncertainty on the numerical evaluations and semi-analytical expressions are two orders of
magnitude larger than claimed. As a result various quantities computed from the size distribution need to be
reevaluated, for instance the resulting relative error on the topological susceptibility at arbitrarily high 
temperatures is about 5\% for QCD and about 10\% for $SU(3)$ Yang-Mills theory. 
With higher rank gauge groups this discrepancy is even higher. We also provide a simple semi-analytical 
formula for the size distribution with absolute error $2\cdot10^{-4}$. In addition we also correct the over-all constant of the
instanton size distribution in the $\msbar$ scheme which was widely used incorrectly in the literature
if non-trivial fermion content is present.
}

\begin{document}

\maketitle

\section{Introduction}
\label{introduction}

This paper is concerned with the semi-classical study of gauge theories. At zero temperature or large
space time volume the semi-classical picture is not reliable but at high temperature and finite spatial volume or small space time
volume, the femto world \cite{Luscher:1981zf, Luscher:1982ma, Koller:1985mb, Koller:1987fq, vanBaal:1988va, vanBaal:1988qm}, it is
because the renormalized coupling runs with the relevant scale, $\mu \sim T$ or $\mu \sim
1/L$, and becomes small. In these regimes the path integral can accurately be computed as the saddle points associated
with instantons and the perturbative fluctuations around them. At asymptotically large temperatures or asymptotically
small space time volumes the higher charge sectors are suppressed relative to lower ones. If an observable only receives
contributions from the non-zero sectors then the leading contribution is from the 1-instanton sector. For instance the
topological susceptibility is such a quantity and the leading semi-classical result is fully determined 
by the size distribution of
1-instantons. This size distribution at high temperature is the object of study in the present paper.

There has been renewed interest in semi-classical results for the topological susceptibility in QCD at high temperature
because of applications in axion physics; see \cite{Wantz:2009it} and references therein, but also \cite{Pisarski:2019upw}  for unrelated
recent developments. Basically, the topological
susceptibility at high temperature can be used to constrain the amount of axions as dark matter components and its mass.
In order to have results from first principles lattice calculations are ideal. Unfortunately the topological susceptibility
decreases fast for $T>T_c$ hence a Monte-Carlo simulation will have a hard time achieving high precision because only very few
configurations fall into the non-zero sectors. Even pure Yang-Mills theory is challenging in this respect
\cite{Berkowitz:2015aua, Borsanyi:2015cka, Frison:2016vuc, Jahn:2018dke}. Beyond pure Yang-Mills, results with dynamical
fermions are also available
\cite{Bonati:2015vqz, Petreczky:2016vrs, Borsanyi:2016ksw, Burger:2018fvb} with full QCD \cite{Borsanyi:2016ksw}
reaching temperatures up to $T/T_c \sim 13-14$.

It is useful to compare the non-perturbative lattice results both in pure Yang-Mills and in full QCD with the
semi-classical results. The relevant semi-classical expressions at finite temperature were first obtained in 
\cite{Pisarski:1980md, Gross:1980br}. The instanton
size distribution consists of two parts, the zero temperature expression and an extra factor responsible for the
temperature dependence. In this work we report the correct zero temperature expression in the $\msbar$-scheme which was 
frequently used incorrectly once light fermions were included and secondly we also report that the factor 
responsible for the temperature
dependence which was numerically evaluated in \cite{Gross:1980br} has an uncertainty that is two orders of magnitude larger
than claimed. In pure Yang-Mills theory only the latter issue is relevant and leads to an increase in the semi-classical 
prediction. Once fermions are included as in QCD, both issues become relevant and while the latter still leads to an
increase, the former leads to a similar decrease.

The organization of the paper is as follows. In section \ref{semiclassical} the basics of the semi-classical expansion
as it relates to high temperatures is summarized, in section \ref{schemedependence} the instanton size distribution is
converted to the $\msbar$ scheme and a frequent error is identified in the literature if non-trivial fermion content is
present. The main result of the paper is contained in \ref{calculationofalambda} where the temperature dependence of the
instanton size distribution is calculated to high precision and in section \ref{topologicalsusceptibility} the results
are applied to the topological susceptibility. Finally we end with some comments and conclusions in section \ref{conclusion}.

\section{Semi-classical expansion at high temperature}
\label{semiclassical}

We will consider $SU(N)$ gauge theory with $N_f$ flavors of light fermions in the fundamental representation.
The semi-classical approximation provides a consistent and reliable expansion of the path integral at fixed spatial
volume and asymptotically 
high temperatures similarly to the situation in small 4-volume or femto world 
\cite{Luscher:1981zf, Luscher:1982ma, Koller:1985mb, Koller:1987fq, vanBaal:1988va, vanBaal:1988qm}. 

It is useful to consider the theory at non-zero
$\vartheta$ angle and the partition function is then given by
\bea
Z(\vartheta) = \sum_{Q=-\infty}^\infty e^{iQ\vartheta} Z_Q\;,
\eea
where $Z_Q$ is the result of integrating over gauge fields with given topological charge $Q$. At
high temperatures the $Q>1$ sectors are suppressed relative to $Q=1$ both exponentially in the inverse coupling as well
as algebraically in the inverse temperature, similarly to the femto world. Hence we have the topological susceptibility
as
\bea
\label{chi2}
\chi = \frac{\langle Q^2 \rangle}{V} = \frac{1}{V} 
\frac{ 2 \left( \frac{Z_1}{Z_0} + 4 \frac{Z_2}{Z_0} + 9 \frac{Z_3}{Z_0} + \ldots \right) }
{1 + 2\left(\frac{Z_1}{Z_0} + \frac{Z_2}{Z_0} + \frac{Z_3}{Z_0} + \ldots \right) } = \frac{2}{V} \frac{Z_1}{Z_0} + \ldots\;,
\eea
where $V = L^3 / T$ is the space-time volume and $L^3$ is the finite spatial box. Note that here we envisage a finite 
and large spatial volume and the limit $T\to\infty$. In this case all further terms in (\ref{chi2}) are 
indeed negligible. For instance there is no need to invoke the dilute instanton gas approximation 
in order to obtain (\ref{chi2}) at $T\to\infty$. 
The subject of the present paper is $Z_1 / Z_0$ and our only objective is to study $Z(\vartheta)$ for $T\to\infty$ 
so there is no reliance on the dilute instanton gas picture at all.
In order to have a self contained presentation and clarify its relationship with the general semi-classical picture,
we will nonetheless summarize the dilute instanton gas model in appendix \ref{diluteinstantongasmodel}. 

Now $Z_1/Z_0$ can be reliably calculated at asymptotically high
temperatures as the contribution of the 1-instanton together with the perturbative fluctuations around it. The leading
term is obtained by
integration over the moduli space of 1-instanton solutions: $(z_\mu, \rho)$ where $z_\mu$ determines
the position and $\rho$ the scale of the instanton. An integration over the arbitrary gauge orientation of the instanton
is implied leading to an $N$-dependent over-all factor. Integration over the position gives a
space-time volume factor $V$.

What we will be concerned with is the integration over the size $\rho$ or 
more precisely the size distribution $n(\rho,T)$,
\bea
\label{topsusc1}
\frac{Z_1}{Z_0} = V \int_0^\infty d\rho n(\rho,T)\;.
\eea
The size distribution at finite temperature \cite{Pisarski:1980md, Gross:1980br} 
can be expressed by the analogous quantity at $T=0$ together with an
explicitly temperature dependent factor $S(\rho,T)$,
\bea
\label{topsusc2}
n(\rho,T) = n(\rho) e^{-S(\rho,T)}\;.
\eea
The zero temperature size distribution $n(\rho)$ with $N_f$ light fermions
at leading order is \cite{tHooft:1976snw, Bernard:1979qt}
\bea
\label{topsusc3}
n(\rho) = C \; \left( \frac{16\pi^2}{g^2(\mu)} \right)^{2N} e^{- \frac{8\pi^2}{g^2(\mu)}} \; \frac{1}{\rho^5} \; ( \rho \mu )^{\beta_1} 
\prod_{i=1}^{N_f} ( \rho m_i(\mu) )\;,
\eea
with $\beta_1 = 11/3 N - 2/3 N_f$ the first coefficient of the $\beta$-function and $C$ is a scheme-dependent constant. 
Its value will be detailed in the next section.
The 2-loop expression for $n(\rho)$ is actually available 
\cite{Morris:1984zi} but for our purposes the leading order result is sufficient. 
Note that the constant $C$ in principle depends on $\rho m_i$ given
by the contribution of the non-zero mode fermion determinant 
\cite{Carlitz:1978yj, Novikov:1983gd, Kwon:2000kf, Dunne:2004sx} but the massless limit can be taken 
for the asymptotically high temperature limit.
It is assumed that the theory is renormalized at $T=0$ by introducing a
renormalized coupling $g^2(\mu)$ and renormalized masses $m_i(\mu)$ at some renormalization scale $\mu$ in a chosen
scheme. Once this is done $n(\rho)$ is finite.

The presence of
a finite temperature does not introduce any new divergences hence $S(\rho,T)$ is also finite and by construction is zero
at $T=0$. 
For definiteness the renormalization scale can be chosen to be $\mu = c T$ with some $O(1)$ constant $c$ and the
dependence on $c$ is indicative of higher order corrections.
Since $S(\rho,T)$ is dimensionless the only dependence is in fact through the variable $\lambda = \pi \rho T$. We
have the result \cite{Brown:1978yj,Gross:1980br},
\bea
\label{topsusc4}
S(\lambda) = \frac{1}{3} \lambda^2 (2 N + N_f ) + 12 A(\lambda) \left( 1 + \frac{N - N_f}{6} \right)
\eea
where the function $A(\lambda)$,
\bea
\label{adef}
12 A(\lambda) = \frac{1}{16\pi^2} \left[ \int_{S^1 \times R^3} 
\left( \frac{\partial_\mu \Pi \partial_\mu \Pi}{\Pi^2}\right)^2 - 
\int_{R^4}
\left( \frac{\partial_\mu \Pi_0 \partial_\mu \Pi_0}{\Pi_0^2} \right)^2
\right]
\eea
encodes the temperature dependence of the 1-loop determinant in the background of the instanton. Here $\Pi(\tau,r)$ 
determines the 1-instanton solution at finite temperature, the periodic Harrington-Shepard solution
\cite{Harrington:1978ve}, while $\Pi_0(t,r)$ is the corresponding function at zero temperature \cite{Belavin:1975fg},
\bea
\Pi_0(t,r) &=& 1 + \frac{\rho^2}{t^2+r^2} \\
\Pi(\tau,r) &=& 1 + \sum_{n=-\infty}^\infty \left( \Pi_0\left(\tau + \frac{n}{T},r\right) - 1 \right) = 
1 + \frac{\pi \rho^2 T }{r} \frac{\sinh(2\pi r T)}{\cosh(2\pi r T) - \cos( 2\pi \tau T )}\;. \nn
\eea
Both the first and second term in (\ref{adef}) are divergent separately and the second term, formally a constant, 
corresponding to zero temperature is subtracted in order to make $A(\lambda)$ finite 
and $A(0) = 0$.

Summarizing the above, once $A(\lambda)$ is found the high temperature limit of $Z_1 / Z_0$ and consequently that of the
topological susceptibility is known.

The main object of study in the present paper is $A(\lambda)$. The definition (\ref{adef}) can not be evaluated
analytically. Numerically, the task is performing a 2 dimensional integral over $(\tau,r)$ and the careful subtraction
of an infinite constant. This was first attempted in \cite{Gross:1980br} and the numerical result was parametrized as,
\bea
\label{agpy}
12 A_{GPY}(\lambda) = - \log\left( 1 + \frac{\lambda^2}{3} \right) + 
\frac{12 \alpha}{\left( 1 + \delta \lambda^{-3/2} \right)^8}\,,
\eea
with $\alpha = 0.01289764$ and $\delta = 0.15858$ and a quoted absolute numerical uncertainty of at most
$12\cdot5\cdot10^{-5}=6\cdot10^{-4}$.
Note that $\delta$ in \cite{Gross:1980br} was denoted by $\gamma$ but in order not to confuse it with Euler's constant 
later we will label it by $\delta$.
Formula (\ref{agpy}) was then used in all subsequent applications 
\cite{Berkowitz:2015aua, Borsanyi:2015cka, Kitano:2015fla, Frison:2016vuc} where the topological susceptibility was
studied at high temperatures and compared with the semi-classical results. 

In the present paper we show that the actual numerical accuracy of (\ref{agpy}) is two orders of magnitude larger than
claimed in \cite{Gross:1980br} 
at around $\lambda \sim O(1)$ leading to about a 5\% mismatch for the topological susceptibility in QCD
or 10\% mismatch for $SU(3)$ pure Yang-Mills. For higher rank gauge groups the mismatch is even higher. Any absolute
error on $A(\lambda)$ needs to be scaled by $N$ for the absolute accuracy of $S(T,\rho)$ and this scaled absolute accuracy
becomes the relative accuracy of the topological susceptibility. 

Before calculating $A(\lambda)$ accurately in section \ref{calculationofalambda} we make a comment in the next section
on the constant $C$ appearing in
(\ref{topsusc3}). It turns out its flavor number dependence in the $\msbar$ scheme was widely used
incorrectly in the literature, especially recently in axion mass estimates as well as comparisons with lattice results.

\section{Scheme dependence - conversion to $\msbar$}
\label{schemedependence}

Historically, there has been quite a bit of confusion about the 
constant $C$ in (\ref{topsusc3}). The topological susceptibility is
of course a finite RG invariant quantity but its perturbative expansion in terms of a
scheme dependent running coupling involve the scheme dependent constant $C$. In the original paper \cite{tHooft:1976snw} 
a Pauli-Villars regulator was used for $SU(2)$ which was extended to $SU(N)$ still with a Pauli-Villars regulator in
\cite{Bernard:1979qt}. The Pauli-Villars results in \cite{tHooft:1976snw} were correct except for some trivial mistakes
corrected later in an erratum, but the conversion to the MS scheme was incorrect. The correct conversion factor was
later found in \cite{Hasenfratz:1980kn} and confirmed in \cite{tHooft:1986ooh}. It is possible to work directly within
the MS scheme lending further support to the correct conversion factor \cite{Luscher:1982wf}. 
The conversion between various schemes
is via the $\Lambda$-parameter ratios, more precisely the constants $C$ in two different schemes are related by
\bea
\label{scheme1}
C_{1} = C_{2} \left( \frac{\Lambda_2}{\Lambda_1} \right)^{\beta_1}\;.
\eea
For definiteness we record the ones relevant for us below
\cite{Bardeen:1978yd, Hasenfratz:1980kn},
\bea
\label{scheme2}
\frac{\Lambda_{\msbar}}{\Lambda_{\rm MS}} = e^{\frac{1}{2}( \log(4\pi) - \gamma )} \qquad \qquad
\frac{\Lambda_{\rm PV}}{\Lambda_{\rm MS}} = e^{\frac{1}{2}(\log(4\pi) - \gamma) + \frac{1}{22}} \qquad \qquad
\frac{\Lambda_{\rm PV}}{\Lambda_{\msbar}} = e^{\frac{1}{22}}\;.
\eea
Note that wrong $\Lambda$-parameter ratios were published for instance in \cite{Weisz:1980pu} as well as
\cite{Dashen:1980vm}. The $\msbar$ scheme is one of the most frequently used schemes and the 
constant $C$ can trivially be obtained from
either the Pauli-Villars or the MS scheme. The result is
\bea
\label{cmsbar}
C_{\msbar} &=& \frac{e^{c_0 + c_1 N + c_2 N_f }}{(N-1)!(N-2)!} \nn \\
c_0 &=& \frac{5}{6} + \log 2 - 2 \log\pi  = -0.76297926 \\
c_1 &=& 4 \zeta^\prime(-1) + \frac{11}{36} - \frac{11}{3}\log 2 = - 2.89766868 \nn \\
c_2 &=& - 4 \zeta^\prime(-1) - \frac{67}{396} - \frac{1}{3}\log 2 = 0.26144360 \;, \nn
\eea
with the derivative of the Riemann $\zeta$-function $\zeta^\prime(s)$.
Even though there is now consensus on the pure Yang-Mills case, the flavor coefficient $c_2$ in $\msbar$ was frequently
quoted \cite{Ringwald:1998ek} as $2\alpha(1/2) = 0.291746$ in the notation of \cite{tHooft:1976snw} and subsequently this value
was used in applications \cite{Berkowitz:2015aua, Borsanyi:2015cka, Frison:2016vuc}.
\footnote{In an earlier work $0.153$ was reported \cite{Moch:1996bs} based on \cite{Balitsky:1992vs} which is also
incorrect.}  
However $2\alpha(1/2)$  equals $- 4 \zeta^\prime(-1) - \frac{5}{36} - \frac{1}{3}\log 2$ which is the relevant result 
within the Pauli-Villars scheme, the one obtained in \cite{tHooft:1976snw}. 
It appears that only the coefficient $c_1$ was converted to $\msbar$ and the flavor coefficient $c_2$ was not 
although of course $\beta_1$ depends on $N_f$.
The mismatch in $c_2$ is precisely $\frac{1}{33} = \frac{2}{3} \cdot \frac{1}{22}$ which should be included using
(\ref{scheme1}) and (\ref{scheme2}). Here $\frac{2}{3}$ comes from the $\beta$-function and $\frac{1}{22}$ from the
ratio of the $\Lambda$-parameters. 

Hence we believe that the fully correct $\msbar$ expressions for the instanton size distribution 
including fermions are given by (\ref{topsusc3}) and (\ref{cmsbar}).

\section{Calculation of $A(\lambda)$}
\label{calculationofalambda}

It is seemingly a relatively straightforward task to evaluate (\ref{adef}). However analytical evaluation is
apparently impossible and numerical integration is tricky primarily because of the subtraction of the
infinite constant and a need for high arithmetic precision. 

The 2 dimensional integral responsible for the $T$-dependence 
is over $(\tau,r)$ where $\tau$ is a periodic variable with period $1/T$. First, we will
perform this integral over $S^1$ analytically and only the second integral over $r$ will be left numerically. Let us
rescale both $\tau$ and $r$ by $1/(2\pi T)$ so that they become dimensionless and $\tau$ periodic with period $2\pi$.
Then we have
\bea
\left( \frac{ \partial \Pi}{\Pi} \right)^4 &=& 16\lambda^8 
\frac{(a\cos^2\tau + b\cos\tau + c)^2}{(X - \cos\tau )^4 ( Y - \cos\tau )^4}  \\
X = \cosh r\;, \qquad Y &=& \cosh r + 2 \lambda^2 f\;, \qquad  f = \frac{\sinh r}{r} \nn \\
a = {f^\prime}^2 - f^2\;, \quad b &=& - 2 f^\prime ( r a + f^\prime f ) \;, \quad
c = ( r a + f^\prime f )^2 + f^2 \nn
\eea
where prime denotes differentiation with respect to $r$. Now the $\tau$-integral can be done analytically for
instance via the residue theorem and we obtain,
\bea
\label{F}
\frac{1}{2\pi} \int_0^{2\pi} d\tau \left( \frac{\partial \Pi}{\Pi} \right)^4 &=& I(r) = 16\lambda^8 \frac{-1}{36} 
\frac{\partial^3}{\partial X^3} \frac{\partial^3}{\partial Y^3} \frac{F(X) - F(Y)}{X-Y}  \\
F(Z) &=& \frac{(aZ^2 + bZ + c)^2}{\sqrt{Z^2-1}} \nn  
\eea
where the $X$ and $Y$ derivatives should be taken at fix $a,b,c$ and in the end $a,b,c,X,Y$ are all simple functions of
$r$.

In a similar fashion let us integrate out $t$ in the second term of 
(\ref{adef}) containing $\Pi_0$ responsible for subtracting an infinite
constant. In this term $r$ can be made dimensionless by rescaling by $\rho$ and we obtain,
\bea
\label{i0}
\frac{1}{2\pi}\int_{-\infty}^\infty dt \left( \frac{\partial \Pi_0}{\Pi_0} \right)^4 =  
\frac{(r+2(r^2+1)^{1/2}) ( 5r^3 + 15r^2(r^2+1)^{1/2} + 18r(r^2+1) + 4(r^2+1)^{3/2} ) }
{2 r^3 (r^2+1)^{7/2} ( r + (r^2+1)^{1/2} )^5 } = I_0(r)\;. \nn
\eea
Now we have both terms in (\ref{adef}) in terms of a dimensionless $r$ and arrive at,
\bea
\label{finite}
12 A(\lambda) = \frac{1}{4\pi} \int_0^\infty dr r^2 \left[ 
\int_0^{2\pi} d\tau \left( \frac{ \partial \Pi }{ \Pi } \right)^4 - 
\int_{-\infty}^\infty dt \left( \frac{\partial \Pi_0 }{ \Pi_0} \right)^4 
\right] =  \frac{1}{2} \int_0^\infty dr r^2 ( I(r) - I_0(r) )\;.
\eea
The integrals with $r^2 I(r)$ or $r^2 I_0(r)$ are separately
divergent and the divergence comes from the origin. For finite and fixed $\lambda$ 
they both behave as $2/r + O(r)$ for $r \ll 1$ hence the difference is finite. For $r \gg 1$ both terms are integrable
separately. More specifically we have,
\bea
\frac{1}{2} r^2 ( I(r) - I_0(r) ) = \left\{
\begin{array}{cc} 
r \ll 1 \quad & -\frac{4}{3}r + O(r^2) \\
\quad & \quad \\
r \gg 1 \quad & \frac{8\lambda^8}{r^6} + O\left(\frac{1}{r^7}\right)
\end{array}
\right. \;.
\eea
Hence (\ref{finite}) defines a convergent integral for $A(\lambda)$. Numerical evaluation is straightforward although
high arithmetic precision is required because of large cancellations. First, the $O(1/r)$ cancellation between $r^2I(r)$
and $r^2I_0(r)$ in (\ref{finite}) needs to be resolved and also the evaluation of $I(r)$ itself via 
(\ref{F}) involve large cancellations especially for small $\lambda$. As we will see the discrepancy between our result
and \cite{Gross:1980br} is precisely here in the small $\lambda$ region.
We have found that keeping $O(100)$ significant digits is nonetheless sufficient.

In order to have as much analytic control as possible it is useful to derive the small $\lambda$ and large $\lambda$
asymptotics which will be used to check the numerical results.

\subsection{Small and large $\lambda$ asymptotics}

It is relatively straightforward to obtain the $\lambda \to 0$ asymptotics. The simplest is to change the integration
variable $r \to \lambda r$ in (\ref{finite}), 
expand the integrand in $\lambda \to 0$ and perform the $r$-integral. The first 3 terms are,
\bea
\label{asympt1}
12 A(\lambda) = - \frac{1}{3} \lambda^2 + \frac{1}{18} \lambda^4 - \frac{1}{81} \lambda^6 + O(\lambda^7) = 
- \log\left( 1 + \frac{\lambda^2}{3} \right) + O(\lambda^7)
\eea
where the $O(\lambda^7)$ terms may include fractional powers and/or logarithms and it turns out that the log on the
right hand side has the same first 3 Taylor coefficients as the ones obtained from the integral. These 3 terms will be
referred to as LO, NLO, NNLO in the following.

The $\lambda \to \infty$ regime is more difficult to obtain. Starting with the original expression (\ref{finite}) one
may split the integral to two intervals, $(0, r_0)$ and $(r_0 , \infty)$ with some fixed $r_0$. In the
$\lambda\to\infty$ limit the first term is finite whereas the second term is logarithmically divergent.

More specifically, the first term can be
straightforwardly expanded in $\lambda\to\infty$ leading to a constant and subleading $O(1/\lambda)$ expressions. 
In the integral $(r_0, \infty)$ all exponentials $e^{-r}$ may be dropped,
the integration over $r$ can then be performed and the result can be expanded in $\lambda\to\infty$. 
Finally in the sum of the two integrals, order by order in $\lambda$, $r_0\to\infty$ can be taken in 
order to justify the dropping of all exponentials. This procedure leads
to, 
\bea 
\label{asympt2}
12 A(\lambda) &=& - \log ( \lambda^2 ) + C_1 - \frac{\log(\lambda^2)}{\lambda^2} - \frac{C_2}{\lambda^2} +
O\left(\frac{1}{\lambda^3}\right) \nn \\
C_1 &=& 2 \left( \frac{1}{3} - \frac{\pi^2}{36} - \gamma + \log\pi \right)  = 1.25338375 \nn \\
C_2 &=& 1 + \log 2 + \frac{\pi^2}{36} + \gamma - \log\pi  = 1.39978864
\eea 
The divergent $- \log(\lambda^2)$ piece is originating from
the second integral $( r_0,  \infty)$ in the above split. The 4 terms above will be referred to as LO, NLO, NNLO
and N$^3$LO in the following, similarly to the small $\lambda$ expansion.

Clearly, the parametrization (\ref{agpy}) satisfies the small $\lambda$ expansion (\ref{asympt1}). However, even though
the small $\lambda$ expansion of the second term of (\ref{agpy}) is $O(\lambda^{12})$ its coefficient is rather large
$12\alpha/\delta^8 \approx 4\cdot 10^{5}$ and distorts the small $\lambda$ behavior. Quite remarkably
the NLO large $\lambda$ expansion (\ref{asympt2}) matches (\ref{agpy})
since the constant term in (\ref{agpy}) is $\log(3) + 12\alpha$ which agrees
with $C_1$ to 6 significant digits. However the NNLO and N$^3$LO terms of the large $\lambda$ expansions do not match
(\ref{agpy}) but for the application to the topological susceptibility this matters much less than the mismatch at
$\lambda = O(1)$.

\subsection{Numerical results}

\begin{figure}
\begin{center}
\includegraphics[width=7.5cm]{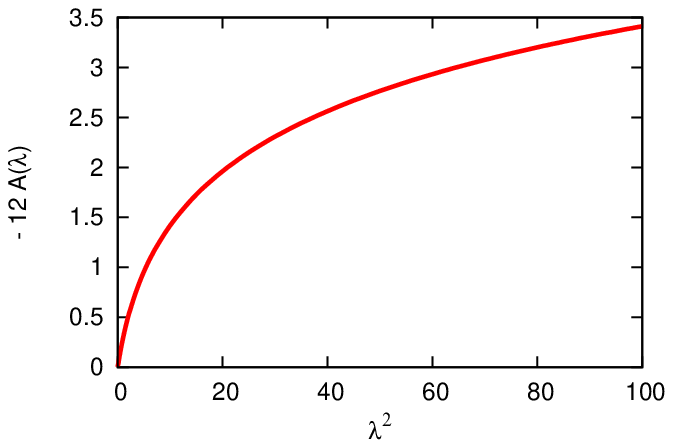} \includegraphics[width=7.5cm]{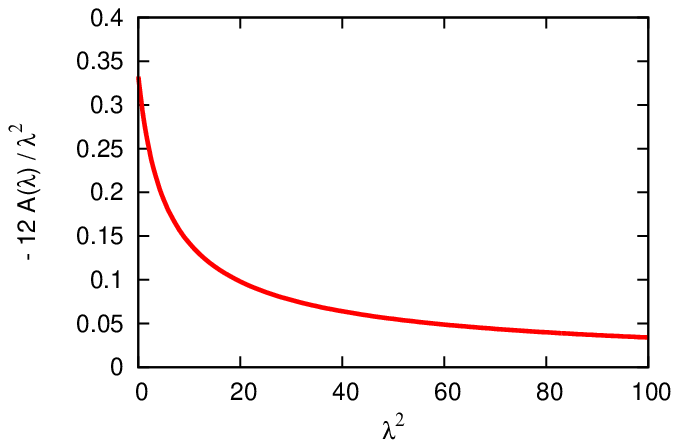}
\end{center}
\caption{The numerically evaluated $A(\lambda)$ using (\ref{finite}).}
\label{aplot}
\end{figure}

After making sure the arithmetic precision is high enough it is straightforward to calculate $A(\lambda)$
numerically using (\ref{finite}). First, numerical integration is done on the interval $(0,8)$ with either the trapezoid
or Simpson's rule with step size $10^{-4}$. The difference between the two schemes is at most $10^{-6}$ which is the
estimated accuracy. The remaining
integral on $(8,\infty)$ can be approximated by dropping all exponentials $e^{-r}$ in the integrand and
performing the integral analytically. This second piece is then added to the numerical integral on the $(0,8)$ finite
interval and amounts to at most $O(10^{-4})$. We have checked that the exact integrand at $r=8$ agrees with the one
where the exponentials are dropped to either $O(10^{-6})$ or the added term is at most $O(10^{-9})$. The procedure is
then reliable in absolute precision to at least $O(10^{-6})$.

\begin{figure}
\begin{center}
\includegraphics[width=7.5cm]{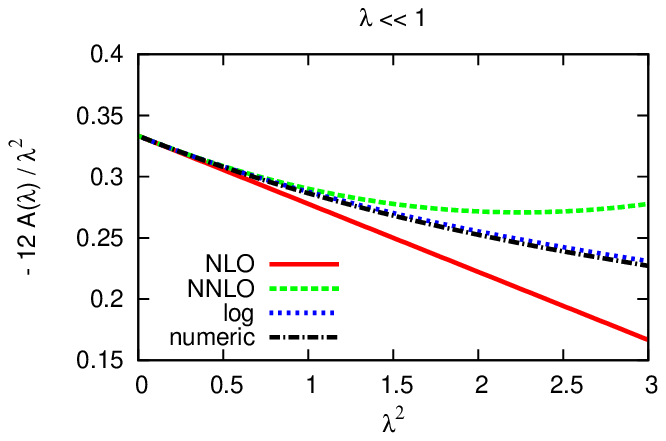} \includegraphics[width=7.5cm]{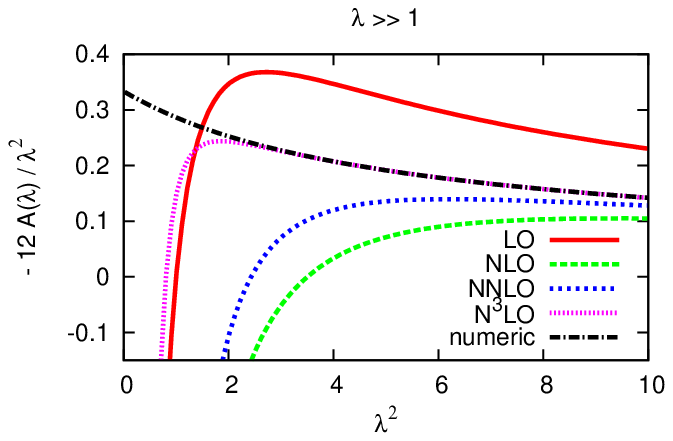}
\end{center}
\caption{Comparison of $-12A(\lambda)/\lambda^2$ with the two asymptotic regimes. 
Left: small $\lambda$ expansion (\ref{asympt1}) with ${\rm log}$ referring to the
right most side of (\ref{asympt1}). Right: the large $\lambda$ expansion (\ref{asympt2}).}
\label{largelambda}
\end{figure}

The final result is shown in figure \ref{aplot}. The numerical result is compared with the
increasing orders of both the small and large $\lambda$ expansions in figure \ref{largelambda} showing increasing
agreement order by order in both regimes as expected.

\begin{figure}
\begin{center}
\includegraphics[width=7.5cm]{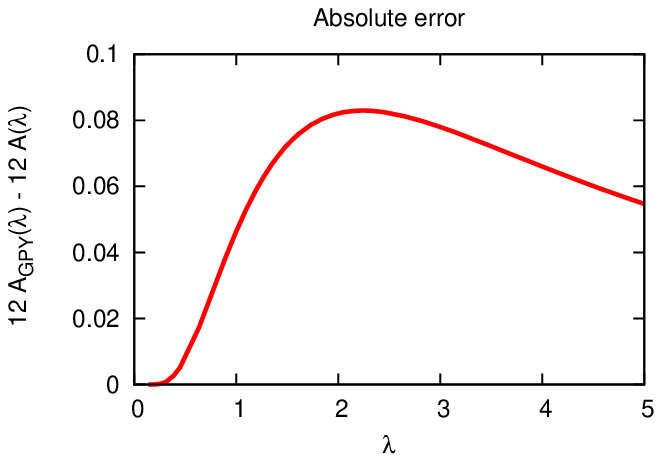} \includegraphics[width=7.5cm]{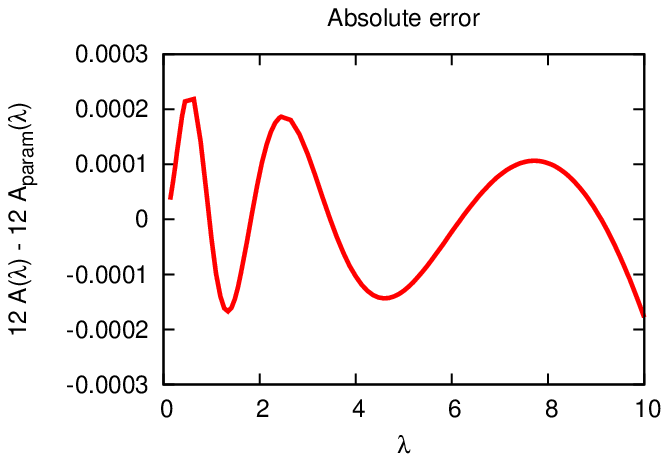}
\end{center}
\caption{Left: the difference between (\ref{agpy}) and our numerical result. Right: the difference between the
parametrization (\ref{param}) and the numerical result.}
\label{diff}
\end{figure}

Let us now compare with \cite{Gross:1980br}. We show the difference between
the parametrization (\ref{agpy}) and the numerical result in the left panel of figure \ref{diff}. 
Clearly, the difference is well outside
the claimed absolute precision of $12\cdot5\cdot10^{-5} = 6\cdot10^{-4}$. 
Note that any absolute error on $A(\lambda)$ will turn into a relative error on
the size distribution. Unfortunately, the differences are largest for $\lambda = O(1)$ which is precisely the region
which contributes the most to the topological susceptibility through the integration over the instanton sizes.

We have found that the following is a convenient parametrization with at most $2\cdot 10^{-4}$ deviation
from the numerical $A(\lambda)$ over the full range $0 < \lambda < 10$,
\bea
\label{param}
-12A_{param}(\lambda) &=& p_0 \log ( 1 + p_1 \lambda^2 + p_2 \lambda^4 + p_3 \lambda^6 + p_4 \lambda^8 ) \\ \nn \\ 
p_0 = 0.247153244,\;\;\;\; p_1 = 1.356391323,\;\;\;\;  p_2 &=& 0.675021523,\;\;\;\; p_3 = 0.145446632,\;\;\;\;  p_4 = 0.008359667 \nn 
\eea
The difference with respect to the numerically evaluated $A(\lambda)$ is shown in the right panel of figure \ref{diff}.  

\section{Topological susceptibility at high temperature}
\label{topologicalsusceptibility}

Once a reliable result for $A(\lambda)$ is obtained the topological susceptibility at asymptotically high temperatures
can be calculated using (\ref{topsusc1}) - (\ref{topsusc4}). Here we are interested in studying the discrepancy between
using (\ref{agpy}) and our result (\ref{param}) for the topological susceptibility for various gauge groups and fermion
content. It is a simple exercise to compute the integrals over the instanton size and it turns out that there is an
approximately 10\% discrepancy between them for $SU(3)$ pure Yang-Mills and about 7\%, 6\% and 4\% for 2, 3 and 4 light flavors,
respectively, with the correct results being larger by these amounts. 
For fixed $N$ the largest relative difference is always at $N_f = 0$ and for increasing $N$ the discrepancy
grows. For instance with $SU(10)$ we have about 22\% while with $SU(20)$ it is about 40\% which is quite sizable. The
reason for this exploding discrepancy is that in the large-$N$ limit the size distribution is more and more peaked
around $\lambda_0 \approx 1.6935$ which unfortunately is approximately where 
the discrepancy between our result and \cite{Gross:1980br} is
largest in absolute terms, see left panel of figure \ref{diff}, and any absolute discrepancy is scaled up by $N$. 
The large absolute discrepancy is translated into a large relative discrepancy for the topological susceptibility. Since
only instantons of the critical size $\lambda_0$ contribute in the large-$N$ limit, a straightforward explicit
expression can in fact be obtained for the topological susceptibility.

Recently there has been renewed interest in the non-perturbative determination of the topological susceptibility in
lattice calculations \cite{DelDebbio:2004vxo, Berkowitz:2015aua, Borsanyi:2015cka, Kitano:2015fla, 
Bonati:2015vqz, Frison:2016vuc, Petreczky:2016vrs, Borsanyi:2016ksw, 
Dine:2017swf, Burger:2018fvb, Jahn:2018dke}
motivated mainly by axion physics. Even the pure Yang-Mills case is challenging and obtaining high precision results at
high enough temperatures is difficult. It is nevertheless clear \cite{Borsanyi:2015cka, Jahn:2018dke} that at
$T/T_c = 2.5$ the temperature is not high enough for the semi-classical topological susceptibility to agree with the
non-perturbative lattice result, however at $T/T_c = 4.1$ the two already only deviate \cite{Jahn:2018dke} within $3\sigma$. The
lattice result \cite{Jahn:2018dke} extrapolated to the continuum is $\log \left( \chi/T_c^4 \right) = -12.47(21)$,
whereas the semi-classical one is $-13.80(10)(40)$ where the 2-loop result for the
susceptibility itself \cite{Morris:1984zi} and 5-loop running \cite{Baikov:2016tgj, Herzog:2017ohr} 
for the coupling was used (however 3-loop or 4-loop running for the coupling gives identical results). 
The first error estimate includes the residual dependence on the scale $\mu$ and the second
(dominant) one originates from the uncertainty of $T_c / \Lambda_{\msbar} = 1.26(7)$ which was also obtained in
lattice calculations \cite{Borsanyi:2012ve}. Note that without lattice input it is only possible to obtain the
semi-classical susceptibility in $\Lambda_{\msbar}$ units with the temperature also measured in $\Lambda_{\msbar}$,
however the lattice results for the susceptibility are in $T_c$ units so the uncertainty related to $T_c /
\Lambda_{\msbar}$ is unavoidable. The resulting uncertainty is rather large because the $\beta_1 - 4 = 7$ power of $T_c
/ \Lambda_{\msbar}$ enters the semi-classical susceptibility.

In QCD the situation \cite{Borsanyi:2016ksw} is similar in that the semi-classical 
susceptibility is in $\Lambda_{\msbar}$ units and with 4
flavors $\Lambda_{\msbar} = 292(16) {\rm MeV}$; the uncertainty is about 5\% \cite{Tanabashi:2018oca}. This uncertainty is
amplified by the even higher factor $\beta_1 - 4 + N_f = 8.33$ in the logarithm of the susceptibility, leading to the
semi-classical result $\log\left( \chi / {\rm MeV\,}^4 \right) = 1.15(3)(46)$ at $T=2000\,{\rm MeV}$ 
where again the inherent uncertainty of the semi-classical result is negligible compared to the one 
associated with the uncertainty of the scale. The corresponding $3+1$ flavor lattice result extrapolated to the
continuum is $3.99(68)$ hence the deviation between the two is about $3.5\sigma$; for higher temperature the deviation
is decreasing \cite{Borsanyi:2016ksw}. The semi-classical result is consistently below the lattice result, similarly to
pure Yang-Mills, and it is most likely that further perturbative $O(g^2)$ corrections to the 2-loop semi-classical topological
susceptibility are responsible for the current $3.5\sigma$ deviation at the relatively high temperature reachable today
by lattice calculations. Note that in QCD the two issues
reported in this paper (more accurate $A(\lambda)$ and correct $C_{\msbar}$) nearly cancel each other in the final result.

\section{Conclusions}
\label{conclusion}

In this paper we have studied the instanton size distribution in gauge theories with light fermions semi-classically. At
large but finite spatial volume and asymptotically high temperatures the contribution of the 1-instanton 
sector is the exact result, there is no need to invoke the dilute instanton gas model.
We were motivated by the recent interest in the semi-classical calculation of the topological susceptibility
at high temperature which in turn was motivated by axion physics. 

It turned out that there were two issues in previous semi-classical calculations which led to incorrect results.
One, even
at zero temperature the over-all constant of the instanton size distribution was not correctly converted to $\msbar$
when $N_f > 0$ and second, the numerically evaluated 1-loop determinant in the instanton background was two orders of magnitude
less precise than originally thought. We have managed to evaluate these determinants to high precision with an absolute
uncertainty of at most $2\cdot10^{-4}$ by doing the periodic integral over the temperature analytically leaving 
only a one dimensional integral numerically which could be handled by standard methods. Although the discrepancy with
previous work is not large the integration over the instanton size is peaked at around the same point where the
discrepancy is largest. Any discrepancy in absolute terms is then converted to a relative discrepancy for the
topological susceptibility which can be 10\% for $SU(3)$ or larger for higher gauge groups.

Once correct semi-classical expressions are obtained a meaningful comparison with the lattice results can be carried out
however the uncertainty of the scale $\Lambda_{\msbar}$ needs to be taken into account. Taking everything into account
in $SU(3)$ pure Yang-Mills the semi-classical and lattice results already at $T/T_c = 4.1$ are within $3\sigma$ 
\cite{Jahn:2018dke} and with
$3+1$ flavor QCD at $T=2000\,{\rm MeV}$ the deviation is below $3.5\sigma$ \cite{Borsanyi:2016ksw}. The
remaining deviation can surely be reduced by simply considering higher temperatures which is however quite demanding on
the lattice or by calculating further perturbative corrections to the semi-classical result.

\section*{Acknowledgments}

We are grateful to Sandor Katz, Tamas Kovacs, Andreas Ringwald 
and Kalman Szabo for insightful discussions and helpful comments on the
manuscript. The work of DN was supported by NKFIH under the grant KKP-126769.

\appendix
\section{The dilute instanton gas model}
\label{diluteinstantongasmodel}

As emphasized throughout the paper the asymptotically high temperature limit at fixed spatial volume
as well as the small space time volume limit are accessible to semi-classical calculations. In these two regimes
the path integral can be computed by the contribution of instantons and the perturbative fluctuations around them.
If a given accuracy is required, only
a finite number of topological sectors need to be considered. If only the leading behavior is required then one needs to
include only the first sector which gives a non-zero result. Apart from being at high temperature or small space time
volume, no further assumptions are needed and the results can be made arbitrarily precise by increasing the perturbative
order at fixed sector and by including more and more sectors. In this sense the semi-classical results are exact in
these two regimes.

The dilute instanton gas model on the other hand is an uncontrolled approximation. 
The assumption is
the presence of {\em some} positively and negatively charged but otherwise indistinguishable and independent objects 
with a constant density. A priori there does not need to
be any reference to weak coupling or semi-classical objects. 
Then the contribution of all sectors will be fixed in terms of a single variable which can be chosen as $\chi$ and we have,
\bea
\label{dig1}
\frac{Z_Q}{Z_0} \sim \sum_{n-m=Q} \frac{1}{n!m!} \left( \frac{V\chi}{2} \right)^{n+m}
\eea
which immediately leads to,
\bea
\label{dig2}
Z_Q &=& Z_0 \frac{I_{Q}\left(V\chi\right)}{I_0\left(V\chi\right)} \nn \\
Z(\vartheta) &=& \frac{Z_0}{I_0\left(V\chi\right)} e^{V\chi \cos\vartheta} = Z(0)
e^{-V\chi( 1 - \cos \vartheta ) }\;,
\eea
where $I_\nu$ is the modified Bessel-function of the first kind. 
The probability to find a gauge field with charge $Q$ is then,
\bea
\label{dig3}
P_Q = \frac{Z_Q}{\sum_{Q^\prime=-\infty}^\infty Z_{Q^\prime} } = e^{-V\chi} I_Q\left(V\chi\right)\;.
\eea
Hence the single parameter $\chi$ determines all $\vartheta$-dependence and the full topological charge distribution.
Unsurprisingly we have,
\bea
\label{chi}
\langle Q^2 \rangle = \sum_Q P_Q Q^2 = V \chi,
\eea
i.e. the single parameter $\chi$ is indeed the topological susceptibility.

We emphasize again that a priori there does not need to be any reference to weak coupling or semi-classical objects at
all in order for (\ref{dig1}) - (\ref{dig3}) to hold with {\em some} parameter $\chi$. Perhaps it would be more accurate to
call these set of assumptions the {\em ideal gas model} rather than the {\em dilute instanton gas model}
in order to make this distinction clear.
This is because the following two questions are independent at a given temperature:
(A) whether the semi-classical expansion at fixed order is accurate or not and (B) whether the
topological charge distribution and the full $\vartheta$-dependence follows 
(\ref{dig1}) - (\ref{dig3}) with a single parameter $\chi$ or not. 
As to question (A) we know that the semi-classical expansion is reliable at finite spatial volume 
and asymptotically high temperatures without any
further assumptions. At asymptotically high temperatures (B) is also true with $\chi$ given by the semi-classical result
(\ref{chi2}) however at asymptotically high temperatures all the additional assumptions (\ref{dig1}) - (\ref{dig3}) are not 
necessary, the semi-classical picture alone is sufficient to calculate anything. 
Question (B) becomes a non-trivial proposition at finite temperature
$T>T_c$ where perhaps it does hold, although there is no a priori reason for it, with {\em some} parameter $\chi$ which
is however different from the semi-classical result (\ref{chi2}). This is precisely what appears to be the case based on
detailed lattice calculations in the pure Yang-Mills case \cite{Bonati:2013tt, Kovacs:2017pox, Vig:2019wei}. Immediately above $T_c$ the
topological charge distribution seems to follow (\ref{dig3}) but with a $\chi$ which is non-perturbative and does not agree
with the semi-classical formula. This comparison is what we would call testing the {\em ideal gas model}.
It would be very interesting to extend these results to full QCD; 
for some recent results see \cite{Bonati:2018blm}.

\end{document}